\documentclass[aps,showpacs,amsfonts,amsmath, amssymb,twocolumn,floatfix,superscriptaddress]{revtex4}

\usepackage[final]{graphicx}
\usepackage{color} 
\usepackage{blindtext}
\usepackage{epstopdf}
\usepackage{units}

\newcommand{\vct}[1]{\mathbf{#1}}

\renewcommand\Re{\operatorname{Re}}
\renewcommand\Im{\operatorname{Im}}
\newcommand\Tr{{\rm Tr}}

\newcommand{\be}{\begin{equation}}
\newcommand{\ee}{\end{equation}}
\allowdisplaybreaks

\DeclareSymbolFont{bbgreek}{U}{bbold}{m}{n}
\DeclareMathSymbol{\bbmu}{\mathbb}{bbgreek}{'26}
\DeclareMathSymbol{\bbeps}{\mathbb}{bbgreek}{'17}

\begin{document}
\title{Anisotropic particles near surfaces: Self-propulsion and friction}

\date{\today}

\author{Boris M\"uller}
\affiliation{4th Institute for Theoretical Physics, Universit\"at Stuttgart, Germany and Max Planck Institute for Intelligent Systems, 70569 Stuttgart, Germany}

\author{Matthias Kr\"uger}
\affiliation{4th Institute for Theoretical Physics, Universit\"at Stuttgart, Germany and Max Planck Institute for Intelligent Systems, 70569 Stuttgart, Germany}

\begin{abstract} 
We theoretically study the phenomenon of self-propulsion through Casimir forces in thermal non-equilibrium. Using fluctuational electrodynamics, we derive a formula for the self-propulsion force for an arbitrary small object in two scenarios, i) for the object being isolated, and ii) for the object being close to a planar surface. In the latter case, the self-propulsion force (i.e., the force parallel to the surface) increases with decreasing distance, i.e., it couples to the near-field. We numerically calculate the lateral force acting on a hot spheroid near a surface and show that it can be as large as the gravitational force, thus being potentially measurable in fly-by experiments. We close by linking our results to well-known relations of linear response theory in fluctuational electrodynamics: Looking at the friction of the anisotropic object for constant velocity, we identify a correction term that is additional to the typically used approach.
\end{abstract}

\maketitle

\nopagebreak
\section{Introduction}\label{sec:Introduction}
The prediction of an attractive force between two uncharged, perfectly reflecting plane-parallel plates embedded in vacuum by H.~B.~G.~Casimir back in 1948 turned out to be a milestone on the way to modern quantum physics \cite{Casimir}. The equilibrium Casimir effect can be equivalently ascribed to quantum zero-point fluctuations of the electromagnetic field, or to charge and current fluctuations in the plates \cite{Lifshitz}. By introducing objects to the quantum vacuum, forces appear due to topological constraints. Casimir's famous formula for the equilibrium energy $E$
\begin{equation}
E=-\frac{\hbar c\pi^2 A}{720 d^3}\;,
\end{equation}
was rederived and verified many times (e.g. see Ref.~\cite{Milonni}). In the present notation $d$ denotes the separation between the plates, $A$ their surface area, $c$ the speed of light, and $\hbar$ the reduced Planck constant, indicating the quantum nature of the Casimir effect. The Casimir effect is relevant on small length scales (e.g. on the submicron scale). 
Shortly after Casimir's breakthrough, the formalism was further developed to be applicable to any kind of dielectric media at finite temperature \cite{Lifshitz}. The rapid development of Casimir physics culminated in the birth of fluctuational electrodynamics in the 1950s \cite{Rytov}.

On the experimental side, scientists were able to quantitatively verify the existence of the theoretically predicted forces in high-precision measurements for the first time around the turn of the millennium \cite{Lamoreaux,Mohideen}. Consequently, many sources of imprecision in force measurements between objects at close proximity were identified and remedied. A few years ago, the attractivity of the Casimir force was reversed by a suitable choice of interacting materials immersed in a fluid \cite{munday2009measured,feiler2008superlubricity,lee2001repulsive,lee2002afm,milling1996direct}. The reversal of the algebraic sign of the force verified the theoretical predictions made by Lifshitz over 50 years ago \cite{doi:10.1080/00018736100101281}.

Recently, situations out of equilibrium have entered the limelight of theory. In this context, phenomena such as vacuum friction or objects at different temperatures have been investigated \cite{polder1971theory,pendry1997shearing,kardar1999friction,messina2011casimir,PhysRevA.84.042102,rodriguez2012fluctuating,rodriguez2013fluctuating,Long_paper,Vlad_linear_response}.  Casimir forces in thermal non-equilibrium have been computed for a variety of different set-ups, e.g. for parallel plates \cite{antezza2008casimir}, deformed plates \cite{Neq_forces2},  between dielectric gratings \cite{guerout2012enhanced,noto2014casimir}, between cylinders \cite{golyk2012casimir}, between a sphere and a plate \cite{Long_paper}, between atoms and surfaces \cite{henkel2002radiation}, between three bodies \cite{messina2014three}, and for inhomogeneous media \cite{polimeridis2015fluctuating}. Also interactions between Brownian charges at different temperatures have been studied \cite{lui2015out}. Moreover, non-equilibrium Casimir forces have been computed in fluid -- or other classical systems \cite{gambassi2009critical,dean2009non,dean2010out,dean2012out,kirkpatrick2013giant,furukawa2013nonequilibrium}.

Generally, in thermal non-equilibrium, forces can be repulsive \cite{antezza2008casimir}, exhibit different power laws \cite{antezza2006casimir}, show stable  points \cite{kruger2011non,bimonte2011dilution} or levitation \cite{Long_paper}. For two spheres with different temperatures, points of self-propelled pairs have been observed \cite{kruger2011non}, where the two identical spheres feel equal forces in the same direction for a specific choice of parameters.

The subject of self-propulsion has become a very popular topic also in fluid systems, where small particles are propelled through different means \cite{romanczuk2012active,elgeti2015physics,tenhagen2014gravitaxis}.

In this paper, we study the Casimir force for anisotropic objects in thermal non-equilibrium focusing first on self-propulsion, employing methods of fluctuational electrodynamics and classical scattering theory. In Sec.~\ref{sec:Non_eq_force_two_objects}, we review the force formulas for two objects in thermal non-equilibrium from Ref.~\cite{Long_paper}. In Sec.~\ref{sec:Self_Prop_Isolation}, we give a compact expression for the  self-propulsion force for a particle {\it in isolation}, also providing a simplified version valid for a small particle. As an example, we explicitly calculate the force for an almost transparent  janus particle. In Sec.~\ref{sec:Arbitrary_Object_in_Front_of_a _Plate}, we add a smooth plate to our set-up and ask for the lateral Casimir force alongside the plate. In particular, we examine the case where the separation $d$ between particle and plate is much smaller than the thermal wavelength $\lambda_T$ (roughly $\unit[8]{\mu m}$ at room temperature), i.e., in the so-called near field limit. Following the derivation of the lateral Casimir force, we explicitly calculate the case of a spheroid in Sec.~\ref{sec:Lateral_force_spheroid}. Finally, in Sec.~\ref{sec:Linear_Response}, we discuss our results from the viewpoint of linear response theory, arguing for an additional term in the friction for an anisotropic particle moving parallel to the surface. Appendices provide technical details and definitions. 

\section{Non-equilibrium force for two objects}\label{sec:Non_eq_force_two_objects}
In this section, in order to keep this article self-contained, we briefly review the formulae and relations for the non-equilibrium Casimir force, closely sticking to Ref.~\cite{Long_paper}. Readers interested only in the new results of this article may skip this section.
 
\begin{figure}[h]
\begin{center}
\includegraphics[width=0.6\linewidth]{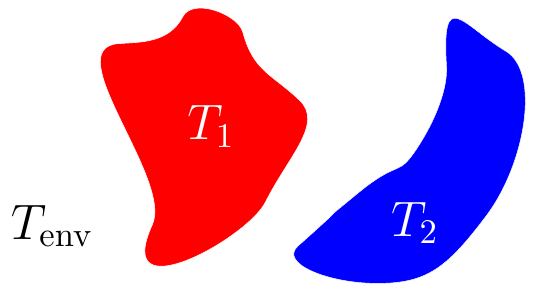}
\caption{Two arbitrary objects at temperatures $T_1$ and $T_2$, respectively, embedded in vacuum in an environment at temperature $T_\mathrm{env}$. Sec.~\ref{sec:Non_eq_force_two_objects} reviews the general force formulas for this situation from Ref.~\cite{Long_paper}.}\label{fig:TwoObjects}
\end{center}
\end{figure}
Let us consider the situation of two arbitrary  (in terms of shape and material properties) objects at different temperatures $T_1$ and $T_2$ embedded in vacuum in an environment at finite temperature $T_\mathrm{env}$ as illustrated in Fig.~\ref{fig:TwoObjects}. In such a non-equilibrium situation the total (Casimir) force acting on object~1 can be written as a sum consisting of all thermal and quantum contributions \cite{Long_paper} 
\begin{align}\label{eq:master}
\vct{F}^{(1)}(T_1, T_2,T_\mathrm{env})=\sum_{\alpha=1,2}\vct{F}^{(1)}_\alpha(T_\alpha)+\vct{F}_\mathrm{env}^{(1)}(T_\mathrm{env})+\vct{F}_0^{(1)}\,.
\end{align}
The terms in the sum, $\vct{F}^{(1)}_1(T_1)$ and $\vct{F}^{(1)}_2(T_2)$  account for the force contributions due to the thermal sources in objects 1 and 2, respectively. $\vct{F}^{(1)}_\mathrm{env}(T_\mathrm{env})$ is the contribution due to thermal fluctuations of the environment. The last term $\vct{F}_0^{(1)}$ incorporates the contribution from zero point fluctuations, i.e., it is the usual zero-temperature Casimir force. By introducing the equilibrium Casimir force $\vct{F}^{(1,\mathrm{eq})}$ at finite temperature, we can get rid of the environment contribution and are able to rewrite the total force as \cite{Correlator_split}
\begin{align}\label{eq:Net_force}
\vct{F}&^{(1)}(T_1, T_2,T_\mathrm{env})\nonumber\\
&=\vct{F}^{(1,\mathrm{eq})}(T_\mathrm{env})+\sum_{\alpha=1,2}[\vct{F}_\alpha^{(1)}(T_\alpha)-\vct{F}^{(1)}_\alpha(T_\mathrm{env})]\,.
\end{align}
This remarkable result states that the force contribution due to sources in the environment does not have to be computed. In the following, we give the different contributions in a basis-independent representation in terms of two well-known quantities: The dyadic free Green's function~$\mathbb{G}_0$ and the classical scattering operator $\mathbb{T}$ for the objects in isolation, both $\mathbb{G}_0$ and $\mathbb{T}$ being  $3\times3$ spacial matrices depending on two position vectors $\vct{r}$ and $\vct{r'}$. The precise definitions of these two quantities are given in Appendix~\ref{app:Greens_function} and \ref{app:Scattering_operator}, respectively.
The first contribution to the force on object~1, $\vct{F}^{(1)}_1(T_1)$ originates from thermal charge and current fluctuations within object 1 itself (which we then refer to as {\it self-force}). In operator notation this term reads \cite{Long_paper}
\begin{align}\label{eq:Self_force}
\vct{F}^{(1)}_1=&\frac{2\hbar}{\pi}\int^\infty_0\mathrm{d}\omega\frac{1}{e^{\frac{\hbar\omega}{k_\mathrm{B}T_1}}-1}\notag\times\\&\Re\Tr\bigg\{\nabla(1+\mathbb{G}_0\mathbb{T}_2)\frac{1}{1-\mathbb{G}_0\mathbb{T}_1\mathbb{G}_0\mathbb{T}_2}\mathbb{G}_0\notag\times\\&[\Im[\mathbb{T}_1]-\mathbb{T}_1\Im[\mathbb{G}_0]\mathbb{T}_1^*]\frac{1}{1-\mathbb{G}_0^*\mathbb{T}_2^*\mathbb{G}_0^*\mathbb{T}_1^*}\bigg\}\,.
\end{align}
Note that each operator product in Eq.~\eqref{eq:Self_force} contains a matrix multiplication as well as a spacial integral over a common coodinate.   The trace in Eq~\eqref{eq:Self_force} finally is meant over both the $3\times3$ matrix as well as the positions $\vct{r}$ and $\vct{r}'$ of the resulting operator. This operator trace can be converted into a more familiar trace over matrix elements in a partial wave representation, yielding closed form equations for specific geometries. For the second contribution to the total force on object 1 evoked by the fluctuations within object $2$ (the {\it interaction force}) one writes \cite{Long_paper}
\begin{align}\label{eq:Interaction_force}
\vct{F}^{(1)}_2=&\frac{2\hbar}{\pi}\int^\infty_0\mathrm{d}\omega \frac{1}{e^{\frac{\hbar\omega}{k_B T_2}}-1}\notag\times\\&\Re\Tr\bigg\{\nabla(1+\mathbb{G}_0\mathbb{T}_1)\frac{1}{1-\mathbb{G}_0\mathbb{T}_2\mathbb{G}_0\mathbb{T}_1}\mathbb{G}_0\notag\times\\&[\Im[\mathbb{T}_2]-\mathbb{T}_2\Im[\mathbb{G}_0]\mathbb{T}_2^*]\mathbb{G}_0^*\frac{1}{1-\mathbb{T}_1^*\mathbb{G}_0^*\mathbb{T}_2^*\mathbb{G}_0^*}\mathbb{T}_1^*\bigg\}\,.
\end{align}

Finally, for completeness, we provide also the more familiar expression for the equilibrium force acting on object~1 at thermal equilibrium at temperature $T$, given by (see, e.g., \cite{Partial_wave_Expansion,Long_paper})
\begin{align}
\vct{F}^{(1,\mathrm{eq})}=&\frac{2\hbar}{\pi}\int_0^\infty \mathrm{d}\omega \bigg[\frac{1}{e^{\frac{\hbar\omega}{k_B T}}-1}+\frac{1}{2}\bigg]\nonumber\times\\
					       &\Im\Tr\bigg\{\nabla\mathbb{G}_0\mathbb{T}_2\frac{1}{1-\mathbb{G}_0\mathbb{T}_1\mathbb{G}_0\mathbb{T}_2}\mathbb{G}_0\mathbb{T}_1\bigg\}\,.\label{eq:eq}
\end{align}
We note, that in contrast to the non-equilibrium force in Eqs.~\eqref{eq:Self_force} and \eqref{eq:Interaction_force} the equilibrium force does not exhibit $\mathbb{G}_0$ being sandwiched by $\mathbb{T}$ operators of the same object \cite{Partial_wave_Expansion}. Besides, the equilibrium force satisfies $\vct{F}^\mathrm{(1,eq)}=-\vct{F}^\mathrm{(2,eq)}$ as expected, while the forces in non-equilibrium are not equal and opposite in general \cite {kruger2011non}. Finally, we recall that Eqs.~\eqref{eq:Self_force} and \eqref{eq:Interaction_force} cannot obviously be integrated to obtain an energy, also in contrast to Eq.~\eqref{eq:eq}. 

In the following section, Sec.~\ref{sec:Self_Prop_Isolation}, we will analyze equation \eqref{eq:Self_force} for the case of one object in isolation, being at a different temperature than the environment (this force is then denoted the {\it self-propulsion} force). In Sec.~\ref{sec:Arbitrary_Object_in_Front_of_a _Plate}, a planar surface is added as object 2, and the force in Eq.~\eqref{eq:Self_force} parallel to the surface is studied (i.e., the change of self-propulsion due to the presence of the surface). 
\section{Self-propulsion for one object in isolation}\label{sec:Self_Prop_Isolation}
\subsection{General expression} 
In order to obtain the total Casimir force for an object in isolation (which we call the {\it self-propulsion force}), we start by removing the second object from Eq.~\eqref{eq:Self_force}, i.e., we set $\mathbb{T}_2=0$, and obtain
\begin{align}\label{eq:Self_force_1}
\vct{F}^{(1)}_1=&\frac{2\hbar}{\pi}\int^\infty_0\mathrm{d}\omega\frac{1}{e^{\frac{\hbar\omega}{k_\mathrm{B}T_1}}-1}\Re\Tr\big\{\nabla\mathbb{G}_0\big[\Im[\mathbb{T}_1]\notag\\&-\mathbb{T}_1\Im[\mathbb{G}_0]\mathbb{T}_1^*\big]\big\}\,.
\end{align}
Due to the facts that $\mathbb{G}_0$ is translationally invariant, $\mathbb{G}_0=\mathbb{G}_0(\vct{r}-\vct{r}')$, and $\mathbb{T}_1$ is a symmetric operator, a partial integration of the first term shows that it is identically zero. It can thus be exactly rewritten to% We thus have the exact expression for the force on the object, i.e., the self-propulsion force $\vct{F_\mathrm{sp}}\equiv\vct{F}^{(1)}_1$,
\begin{align}\label{eq:Self_force_2}
	\vct{F}^{(1)}_1=-\frac{2\hbar}{\pi}\int^\infty_0\mathrm{d}\omega\frac{1}{e^{\frac{\hbar\omega}{k_\mathrm{B}T_1}}-1}\Re\Tr\big\{\nabla\mathbb{G}_0\big[\mathbb{T}_1\Im[\mathbb{G}_0]\mathbb{T}_1^*\big]\big\}\,.
\end{align}
This is the force acting on an isolated arbitrary object at temperature $T_1$ in an environment at zero temperature. In order to obtain the force for a finite $T_\mathrm{env}$, the same expression, evaluated at $T_\mathrm{env}$, must be subtracted (see Eq.~\eqref{eq:Net_force} and recall that the force is zero in equilibrium). We thus have the exact expression for the force on the object, i.e., the self-propulsion force 
\begin{align}\label{eq:Self_force_3}
\vct{F_\mathrm{sp}}(T_1,T_{\rm env})	=&-\frac{2\hbar}{\pi}\int^\infty_0\hspace*{-0.017\linewidth}\mathrm{d}\omega\bigg[\frac{1}{e^{\frac{\hbar\omega}{k_\mathrm{B}T_1}}-1}-\frac{1}{e^{\frac{\hbar\omega}{k_\mathrm{B}T_\mathrm{env}}}-1}\bigg]\notag\times\\&\Re\Tr\bigg\{\nabla\mathbb{G}_0\big[\mathbb{T}_1\Im[\mathbb{G}_0]\mathbb{T}_1^*\big]\bigg\}\,.
	\end{align}
	In the following subsections, we will analyze this term for special cases, thereby for brevity of notation dropping the subscript $1$ and setting $T_\mathrm{env}=0$.
	
\subsection{Exact force in the spherical basis}

By using the techniques presented in detail in Ref.~\cite{Long_paper}, we expand the force formulae of the previous section in the spherical wave basis (see Appendix~\ref{app:Spherical_Basis} for details). Applying this procedure to Eq.~\eqref{eq:Self_force_2}, we obtain
\begin{align}\label{eq:spf_isolation}
	\vct{F_\mathrm{sp}}(T,0)=\frac{2\hbar}{\pi}\int^\infty_0\mathrm{d}\omega\frac{1}{e^{\frac{\hbar\omega}{k_\mathrm{B}T}}-1}\Im\Tr\big\{\vct{p}\mathcal{T}\mathcal{T}^\dagger\big\}\,.
\end{align}
In this equation $\mathcal{T}\equiv\mathcal{T}_{lm,l'm'}^{P,P'}$ comprises the discrete matrix of the classical scattering operator $\mathbb{T}$ for waves with quantum numbers $l$ and $m$ and polarization $P$,  see Appendix~\ref{app:T_Elements}. Primed indices denote incoming waves, while unprimed denote scattered ones.  $\vct{p}$ is the infinitesimal translation operator which plays the role of a spatial derivative \cite{Long_paper}, it is given explicitly in Eq.~\eqref{eq:translation_matrix} below. Finally, the representation of the self-propulsion force in the equation above implies matrix multiplications over the given indices $\{P,l,m\}$. The properties of the spherical basis are introduced in Appendix \ref{app:Spherical_Basis}. For the $i$th component of the self-propulsion force the trace in Eq.~\eqref{eq:spf_isolation} turns into (using the Einstein summation convention)
\begin{align}\label{eq:spf_isolation_trace}
	\Tr\big\{\vct{p}\mathcal{T}\mathcal{T}^\dagger\big\}_i=p_{i;\,Plm,P'l'm'} \mathcal{T}^{P',P''}_{l'm',l''m''}\mathcal{T}^{*\,P,P''}_{lm,l''m''}\,,
\end{align}
where, regarding --without loss of generality--  the $z$-component of the force, one has
\begin{align}\label{eq:translation_matrix}
	p&_{z;Plm,P'l'm'}=-\frac{\omega}{c}\{i\,(1-\delta_{P'P})\,\delta_{l'l}\,a(l,m)\nonumber\\
	 &+\delta_{P'P}[-b(l,m)\,\delta_{l',l+1}+b(l',m)\,\delta_{l'+1,l}]\}\,\delta_{m'm}\,,
\end{align}
with
\begin{align}
	a(l,m)&=~\frac{m}{l(l+1)}\,,\label{eq:a}\\
	b(l,m)&=\frac{1}{l+1}\sqrt{\frac{l(l+2)(l-m+1)(l+m+1)}{(2l+1)(2l+3)}}\label{eq:b}\,.
\end{align}
The self-propulsion force is expected to vanish for isotropic objects, as can be easily seen for the case of a homogeneous sphere, where the matrix  $\mathcal{T}$ is diagonal (see e.g. Eq.~\eqref{eq:Tm} below). As the $\vct{p}$ matrix has only off-diagonal terms, Eq.~\eqref{eq:spf_isolation} is zero for that case. This observation corresponds to our physical expectation, since an object can only be self-propelled if there is a preferred direction of radiation. 

\subsection{Force for a small object}
Eq.~\eqref{eq:spf_isolation} is valid for an object of any size and shape. In this subsection, we aim to collect the leading terms (leading matrix elements) contributing to the self-propulsion force for a small object. Such leading terms will be dominant if the size of the object (denoting $R$ as the largest dimension of the anisotropic object) is the smallest scale involved, i.e., if $R$ is small compared to the thermal wavelength $\lambda_T=\hbar c/k_BT$ (which is roughly $\unit[8]{\mu m}$ at room temperature), as well as the material skin depth. In lowest order in $R$, we find that the off-diagonal elements $\mathcal{T}^{M,N}_{1m,1m'}$ and $\mathcal{T}^{N,N}_{2m,1m'}$ contribute, and more specifically, for the $z$-component of the force, 
	\begin{align}\label{eq:sp}
	 F_{\mathrm{sp},z}\hspace*{-0.01\linewidth}=&\frac{4\hbar}{\pi c}\int^\infty_0\hspace*{-0.042\linewidth}\mathrm{d}\omega\frac{\omega}{e^{\frac{\hbar\omega}{k_\mathrm{B}T}}-1}\hspace*{-0.018\linewidth}\sum_{\substack{|m|<l\\|m'|<l'}}\hspace*{-0.029\linewidth}\Big(a(1,m)\Re[\mathcal{T}^{M,N}_{1m,1m'}\mathcal{T}^{*\,N,N}_{1m,1m'}]\nonumber\\
	 &-b(1,m)\Im[\mathcal{T}^{N,N}_{2m,1m'}\mathcal{T}^{*\,N,N}_{1m,1m'}]\Big)+\mathcal{O}(R^8).
	\end{align}
As $\mathcal{T}^{M,N}_{1m,1m'}$ and $\mathcal{T}^{N,N}_{2m,1m'}$ are of order $\mathcal{O}(R^4)$ and $\mathcal{T}^{N,N}_{1m,1m'}\propto R^3$, the self-propulsion force is found to be of order $R^7$ for small $R$.

\subsection{Force for a dilute object}\label{subsec:Force_for_a_dilute_object}

\begin{center}
\begin{figure}[h]
\includegraphics[width=0.7\linewidth]{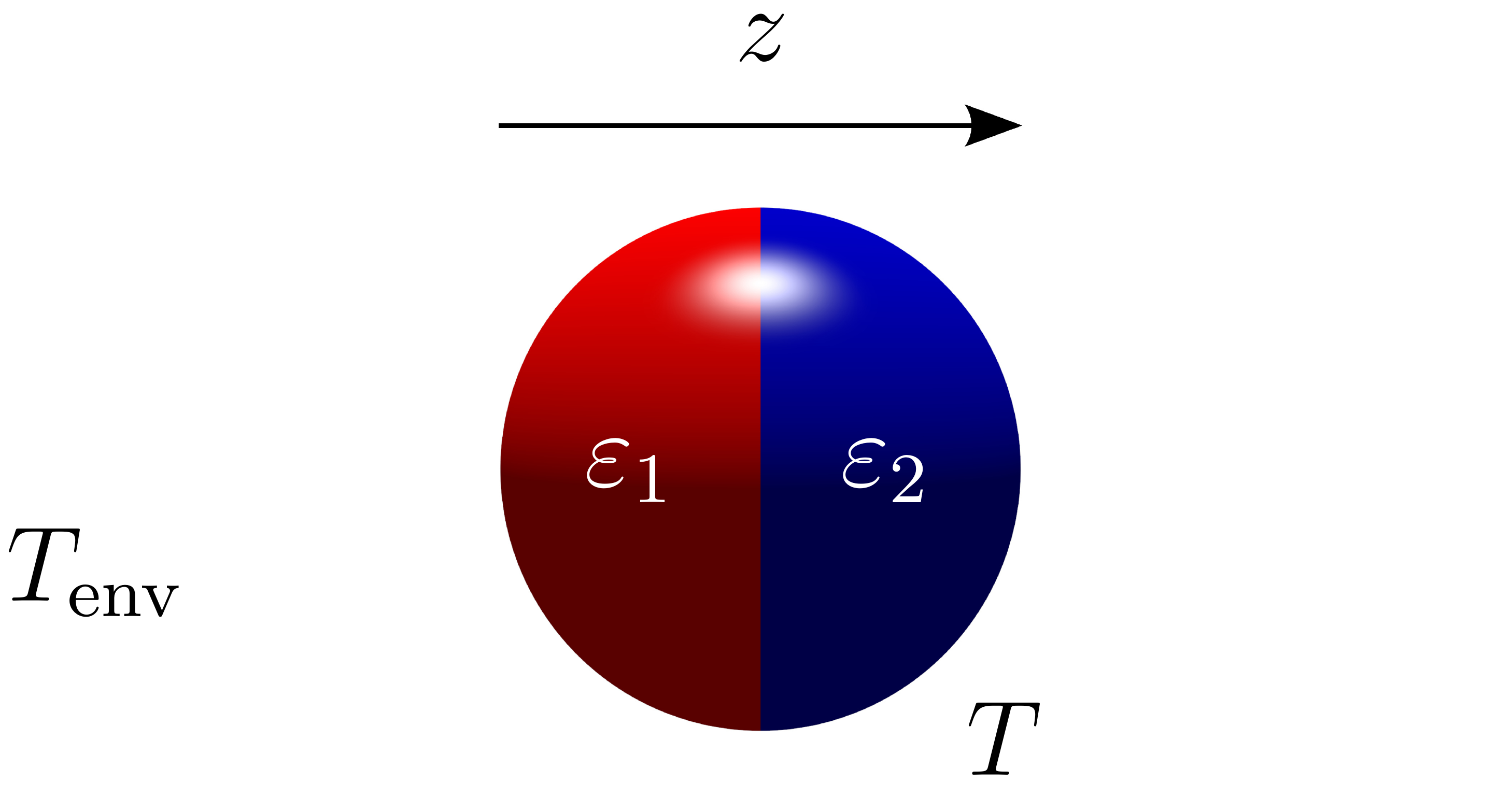}
\caption{A janus particle made up of two hemispheres with different dielectric responses $\varepsilon_1$ and $\varepsilon_2$ at temperature $T$. The particle is embedded in vacuum at temperature $T_\mathrm{env}$.\label{fig:j}}
\end{figure}
\end{center}

In order to demonstrate the power of our derived formulae, we explicitly evaluate the self-propulsion force for a janus particle of radius $R$ at temperature $T$, which is almost transparent,  i.e., $\varepsilon-1\ll 1$, see Fig.~\ref{fig:j}.
In this limit, the classical scattering operator $\mathbb{T}=\mathbb{V}\frac{1}{1-\mathbb{G}_0\mathbb{V}}\approx \mathbb{V}$, where ${\mathbb{V}=\frac{\omega^2}{c^2}(\bbeps-\mathbb{I})+\nabla\times(\mathbb{I}-\frac{1}{\bbmu})\,\nabla\times}$ is the potential introduced by the objects \cite{Partial_wave_Expansion}. Taking $\bbeps$ local and isotropic, the electric response of the janus particle reduces to the scalar ${\varepsilon=[\varepsilon_1\Theta(-z))+\varepsilon_2\Theta(z))]\Theta(R-r)}$, where $\Theta$ is the unit step function and $r$ is the radial distance measured from the center of the sphere. 
By additionally assuming our particle to be non-magnetizable, we can easily calculate the $\mathcal{T}$ matrix elements defined in Appendix \ref{app:T_Elements}. In leading order, we arrive at
\begin{align}\label{eq:Force_janus_particle_isolated}
F_{\mathrm{sp},z}=&\frac{2\hbar}{\pi c^{10}}\int_0^\infty\hspace*{-0.03\linewidth}\mathrm{d}\omega\,\frac{\omega^{10}}{e^{\frac{\hbar\omega}{k_\mathrm{B}T}}-1}\Bigg[\frac{1}{2700}R^9\{\Im[\varepsilon_2](\Re[\varepsilon_1]-1)\nonumber\\
&-\Im[\varepsilon_1](\Re[\varepsilon_2]-1)\}\Bigg]+\mathcal{O}(R^{11})+\mathcal{O}(\varepsilon-1)^3\,.
\end{align}
This result holds up to order $(\varepsilon-1)^3$ and $R^{11}$.
By setting $\varepsilon_1=\varepsilon_2$ we can verify again that the force vanishes for a homogeneous sphere to the given order.

\section{Lateral force on an arbitrarily shaped object in front of a plate in near field limit}\label{sec:Arbitrary_Object_in_Front_of_a _Plate}
\subsection{Introduction}
\begin{figure}[h]
\begin{center}
\includegraphics[width=\linewidth]{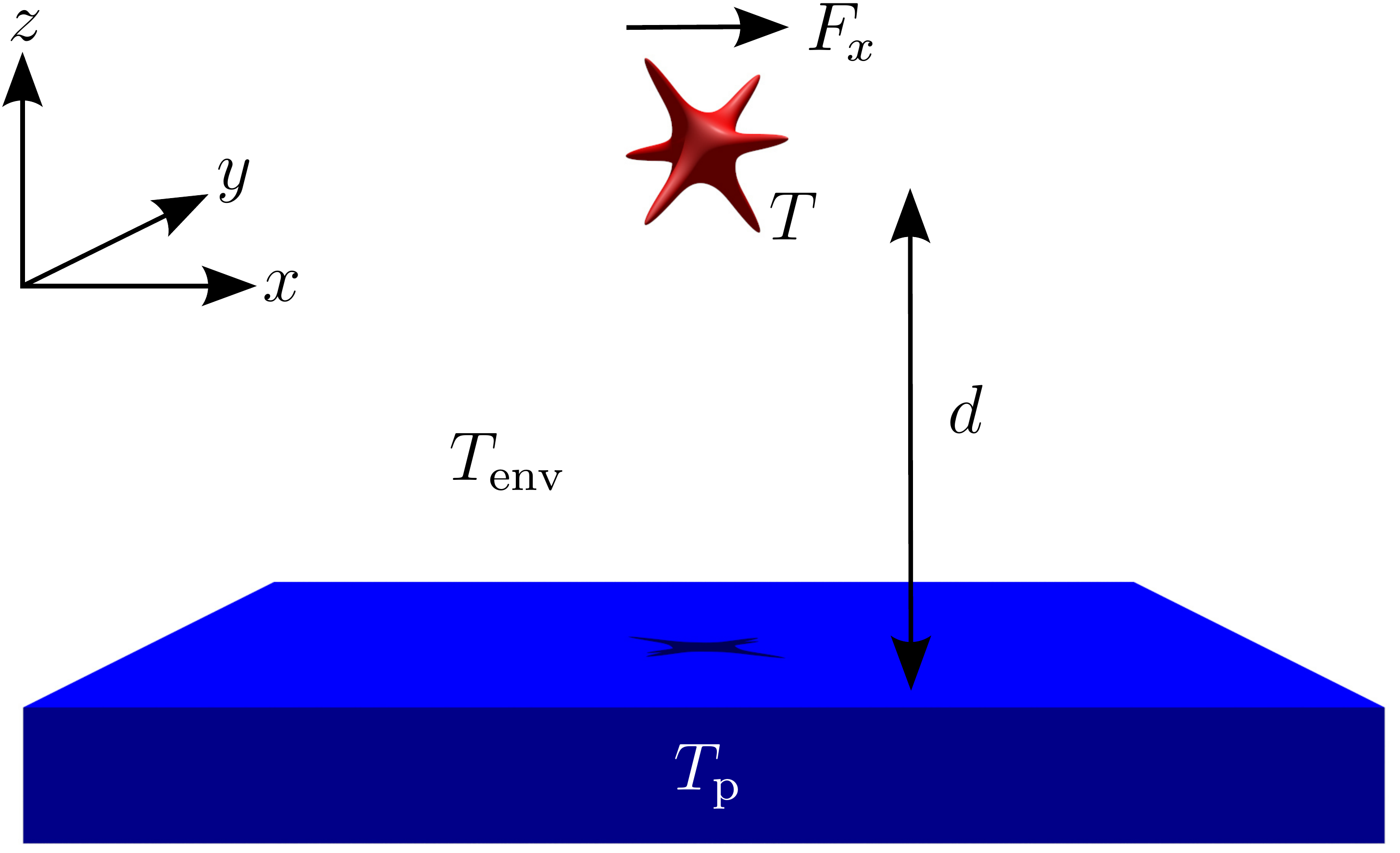}
\caption{The system of an anisotropic particle in front of a plate in thermal non-equilibrium. We compute the lateral Casimir force ($F_x$), which due to translational invariance vanishes identically in equilibrium.}\label{fig:2}
\end{center}
\end{figure}

In a process of miniaturization, manufacturers of technical devices try to decrease the dimension of their products. These compact technical solutions use components with a size of several micrometers or even nanotechnology. On these scales, Casimir forces have to be considered in order to avoid unintended effects such as ``stiction'', meaning components sticking together and disturbing the functionality of the device \cite{Stiction}. For instance, micro-electromechanical systems (MEMS) are hugely influenced in their behavior by Casimir physics. These devices are made up of components between $1$ to $\unit[100]{\mu m}$ in size being located in close proximity. Regarding situations out of equilibrium, it has been found that the forces between a sphere and a plate can show properties very different from equilibrium counterparts, including e.g. levitation \cite{trondle2010critical,Long_paper}. In Ref.~\cite{Long_paper}, the force normal to the surface was investigated. In this section, we want to focus on the other component, i.e.,  pointing {\it alongside} the plate: The lateral Casimir force, which again, we denote the {\it self-propulsion force}, as it propels the object parallel to the surface, i.e., in a direction where the system is translationally invariant, see Fig.~\ref{fig:2}.

In thermal equilibrium, a lateral force cannot be observed for any kind of object due to the mentioned translation invariance of the arrangement. However, in a non-equilibrium situation we cannot invoke such arguments, and, depending on the symmetries of the object, we will indeed observe a lateral force below.
\subsection{Force in the near field limit in terms of $\mathcal{T}$-matrix}
We start again from the exact expression for the self-force in Eq.~\eqref{eq:Self_force}, which, written in the spherical basis reads \cite{Long_paper},
	 \begin{align}\label{eq:Self_force_expanded}
		&F_{1,x}^{(1)}=\frac{2\hbar}{\pi}\int^\infty_0\mathrm{d}\omega\frac{1}{e^{\frac{\hbar\omega}{k_\mathrm{B}T}}-1}\Im\Tr\bigg\{[p_x\mathcal{U}\mathcal{T}_p\mathcal{U}+p_x]\nonumber\times\\
		 &(1+\mathcal{T}\mathcal{U}\mathcal{T}_p\mathcal{U})^{-1}\bigg[\frac{\mathcal{T}^\dagger+\mathcal{T}}{2}+\mathcal{T}\mathcal{T}^\dagger\bigg](1+\mathcal{U}^\dagger\mathcal{T}_p^\dagger\mathcal{U}^\dagger\mathcal{T}^\dagger)^{-1}\bigg\}.
	 \end{align}
  
 We aim to analyze this equation for the case of a small object (with scattering matrix $\mathcal{T}$) in front of a planar surface (with scattering matrix $\mathcal{T}_p$), hence expanding in powers of $\mathcal{T}$ (see App.~\ref{app:Plane_Wave_Basis} for details on the plane wave basis, and App.~\ref{app:Conversion_Matrix} for details on the conversion matrices $\mathcal{U}$).
 Furthermore, we aim at the behavior of the force at small distance $d$, i.e., in the near-field regime with $d\ll\lambda_T$ (the size $R$ of the object is nevertheless assumed small compared to $d$). Interestingly, the leading term, which is linear in both $\mathcal{T}_p$ and $\mathcal{T}$ (the ''one-reflection approximation``) is identically zero due to the translational invariance of the planar surface along $x$. The leading term in the near-field (see Eq.~\eqref{eq:Near_field_limit_T} below) is hence quadratic in both $\mathcal{T}_p$ and $\mathcal{T}$, resulting from two ''reflections``. 

${\cal T}_p$ of the plate turns into the Fresnel coefficients (see Appendix~\ref{app:T_Elements} for details). In the considered limit of $d\ll\lambda_T$, only the Fresnel coefficient for electric polarization contributes, and only its limit for infinite wave vector (where $\varepsilon_p$ is the dielectric function of the plate),
\begin{equation}
r^N(\omega) =\frac{\varepsilon_p(\omega)-1}{\varepsilon_p(\omega)+1}+\mathcal{O}\left(\frac{1}{k_\perp^2}\right)\,.
\end{equation}
We arrive at the following expression for the force, valid for $R\ll d\ll\lambda_T$, (here, the terms with $l=1$ and $P=N$ dominate, and for brevity, we have omitted the index $l$ and superscript $P$ at $\cal T$ and give  only the indices $m$ and $m'$, e.g. $\mathcal{T}_{m',m}=\mathcal{T}^{N,N}_{1m',1m}$)
\begin{align}\label{eq:Near_field_limit_T}
  &\lim_{\lambda_T\gg d\gg R} F_{1,x}^{(1)}(T) =\frac{27}{128\sqrt{2}\pi}\frac{\hbar}{d^7}\int_{0}^\infty \mathrm{d}\omega \frac{1}{e^{\frac{\hbar\omega}{k_BT}}-1}\frac{c^6}{\omega^6}\nonumber\times\\
  &\Im\Bigg[\frac{\varepsilon_p-1}{\varepsilon_p+1}\Bigg]^2\sum_{m}A_{m}\Re[\mathcal{T}_{0,m}\mathcal{T}^*_{1,m}-\mathcal{T}_{-1,m}\mathcal{T}^*_{0,m}]\,,
\end{align}
where $A_{m}$ takes the values $A_{m}=1$ for $m=\{-1,1\}$ and $A_{m}=2$ for $m=0$.
Eq.~\eqref{eq:Near_field_limit_T} allows the computation of the lateral Casimir force in the near-field limit for an arbitrary object in front of a plate in thermal non-equilibrium and constitutes one of our main results. The leading order term is of order $\mathcal{O}(R^6)$ and behaves  like $d^{-7}$ in the given  limit $\lambda_T\gg d\gg R$.
It is striking that only the imaginary part of the reflection coefficient contributes to the lateral force in the given limit. 
The lateral Casimir force in the near-field is an effect due to evanescent wave contribution. We note that the term in Eq.~\eqref{eq:Near_field_limit_T}  vanishes in the limit where the plate approaches a perfect reflector ($\varepsilon_p\to\infty$).

We also computed the force on the object due to the fluctuations in the plate, i.e., the interaction force $\vct{F}^{(1)}_2$ in Eq.~\eqref{eq:Interaction_force}. We found that it exactly equals Eq.~\eqref{eq:Near_field_limit_T} in the given limit, i.e, 
\begin{equation}\label{eq:fe}
\lim_{R\ll d\ll \lambda_T}F_{1,x}^{(1)}(T)=-F_{2,x}^{(1)}(T)\,.
\end{equation}
In the near field limit,  the temperature of the environment is expected  to be negligible; furthermore, as any equilibrium contribution in Eq.~\eqref{eq:Net_force} vanishes for the considered components, Eq.~\eqref{eq:fe} was expected, and its direct confirmation is a consistency check for our computations.

Hence, in the given limit, the total lateral Casimir force can now be computed for any combinations of the temperature $T$ of the object and the temperature $T_p$ of the plate,
\begin{equation}\label{eq:Te}
\lim_{R\ll d\ll \lambda_T}F_x(T,T_p)
=F_{1,x}^{(1)}(T)-F_{1,x}^{(1)}(T_p)\,.
\end{equation}
This relation remains true for the forces found in Eqs.~\eqref{eq:Self_force_expanded}, \eqref{eq:Near_field_limit_T}, \eqref{eq:Lateral_force_general} and \eqref{eq:Lateral_force_spheroid} below.
\subsection{Force in terms of polarizabilites}
Often the polarizability tensor \cite{bohren2008absorption,Tsang_Kong_Ding,landau1984electrodynamics} of small objects (nanoparticles) is better known than the $\cal T$ matrix elements in Eq.~\eqref{eq:Near_field_limit_T}, and we also convert Eq.~\eqref{eq:Near_field_limit_T} into a form containing these explicitly (see Appendix~\ref{app:Polarizability_Tensor} for details). We find 
\begin{align}\label{eq:Lateral_force_general}
	&\lim_{\lambda_T\gg d\gg R} F_{1,x}^{(1)} =\frac{3}{32\pi}\frac{\hbar}{d^7}\int_0^\infty \mathrm{d}\omega
 \frac{1}{\mathrm{e}^{\frac{\hbar\omega}{k_BT}}-1}\nonumber\times\\
 &\Im\left[\frac{\varepsilon_p-1}{\varepsilon_p+1}\right]^2 \Im\left[\alpha_{zx}\alpha_{xx}^*+\alpha_{zy}\alpha_{xy}^*+2\alpha_{zz}\alpha_{xz}^*\right]\,.
\end{align}
Here, $\alpha_{ij}$ is the $ij$ component of the $3\times3$ dimensional polarizability tensor \cite{bohren2008absorption}. The strong dependence of the force on the spatial orientation of the object becomes apparent from Eq.~\eqref{eq:Lateral_force_general}. In the case of a sphere, the off-diagonal components of the polarizability tensor are identically zero and we confirm once more that the lateral force vanishes in that case. 

We finally note that the self-propulsion force for a small object near a surface scales as $R^6$ for small $R$, while it scales as $R^7$ for a free particle, compare Eq.~\eqref{eq:sp}. We also note that the force for an isolated particle cannot be described by a polarizability tensor (higher asymmetries are neccessary) in contrast to Eq.~\eqref{eq:Lateral_force_general}.  Finally, a computation for a dilute janus particle, compare Eq.~\eqref{eq:Force_janus_particle_isolated}, yields in the presence of a surface a force of order $R^7$, so that forces in the presence of a plate are generally much stronger than for isolated particles.

\section{Explicit application: A spheroid in front of a plate}\label{sec:Lateral_force_spheroid}
\subsection{Formula}
\begin{figure}[h]
\begin{center}
\includegraphics[width=\linewidth]{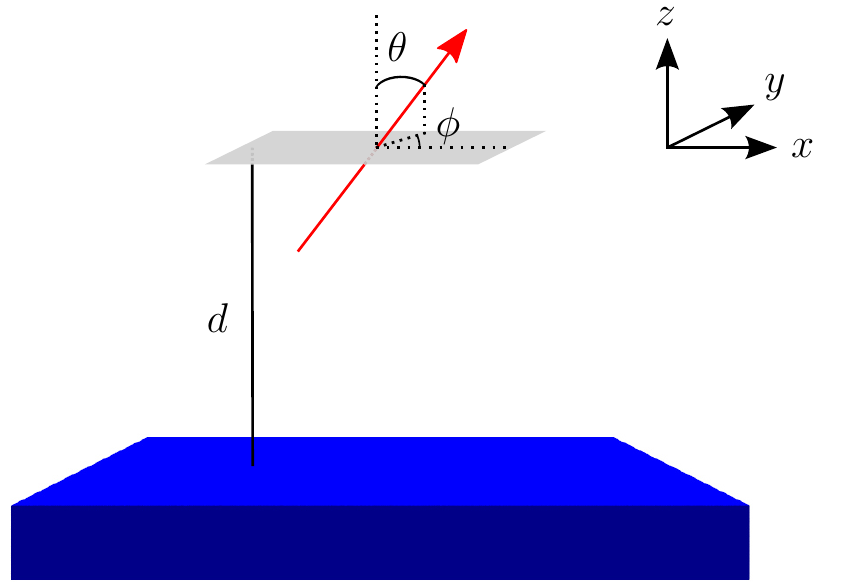}
\caption{\label{fig:Schematic_spheroids}Sketch of an object with axis of rotational symmetry, and the definition of the angles $\theta$ and $\phi$ appearing in Eq.~\eqref{eq:Lateral_force_spheroid}. The object is mimicked by an arrow, which points in the direction of the axis of rotational symmetry.}
\end{center}
\end{figure}

Spheroids, i.e., ellipsoids with an axis of rotational symmetry, are suitable candidates in order to observe lateral Casimir forces in front of a plate in thermal non-equilibrium as they exhibit an explicit anisotropy [equilibrium forces involving ellipsoids have been studied in detail \cite{emig2009orientation,kondrat2009critical,biehs2014anisotropy}, which are however irrelevant for our discussion]. In this section, we focus on the case of a prolate (cigar-shaped) spheroid, for which $R_\parallel>R_\perp$, where $R_\parallel$ and $R_\perp$ denote the radius parallel and perpendicular to the axis of symmetry. The orientation of the spheroid with respect to the surface is described by the two angles $\theta$ and $\phi$, see  Fig.~\ref{fig:Schematic_spheroids}. The spheroid's polarizability can then be expressed by the two components,
\begin{equation}\label{eq:Polarizability_spheroid}
 \alpha_\parallel\equiv \hat{\vct{e}}^T_\parallel\cdot \hat{\alpha}\cdot \hat{\vct{e}}_\parallel, \qquad \alpha_\perp\equiv\hat{\vct{e}}^T_\perp\cdot \hat{\alpha}\cdot \hat{\vct{e}}_\perp\,,
\end{equation}
where $\hat{\vct{e}}_\parallel$ and $\hat{\vct{e}}_\perp$ are unit vectors pointing along the axis of rotational symmetry and perpendicular to it, respectively (and the superscript $T$ denotes the transpose of the vector).
Eq.~\eqref{eq:Lateral_force_general} is then directly rewritten to 
\begin{align}\label{eq:Lateral_force_spheroid}
	\lim_{\lambda_T\gg d\gg R}& F_{1,x}^{(1)} =\frac{3}{64}\frac{\hbar}{d^7}\sin(2\theta)\cos(\phi)\nonumber\times\\
	&\int_0^\infty \mathrm{d}\omega
 \frac{1}{\mathrm{e}^{\frac{\hbar\omega}{k_BT}}-1}\Im\left[\frac{\varepsilon_p-1}{\varepsilon_p+1}\right]^2\Im\left[\alpha_\parallel\alpha_\perp^*\right]\,.
\end{align}
In this equation $T$ is the temperature of the spheroid and the temperature of the plate can be added according to Eq.~\eqref{eq:Te}. The rotation angles $\theta$ and $\phi$ appear as they translate the local coordinate system into the global frame as described in Appendix~\ref{app:Polarizability_Tensor}. The force in $x$-direction is maximal for $\phi=0$ and $\theta=\pi/4$. 

\subsection{Numerical evaluation}
In order to evaluate Eq.~\eqref{eq:Lateral_force_spheroid} numerically, we resort to the polarizabilities for spheroids given in Refs.~\cite{bohren2008absorption,Tsang_Kong_Ding,landau1984electrodynamics},
\begin{equation}
 \alpha_{\parallel/\perp}(\omega)=\frac{1}{3}\frac{R_\perp^2 R_\parallel (\varepsilon(\omega)-1)}{(\varepsilon(\omega)-1)n_{\parallel/\perp}(\eta)+1}\,,
\end{equation}
with the dielectric permittivity $\varepsilon(\omega)$ and the geometrical factors $n_{\parallel/\perp}$ for prolate spheroids,
\begin{align}
 n_\parallel(\eta)&=\
 \frac{1-\eta^2}{2\eta^3}\left(\log\left(\frac{1+\eta}{1-\eta}\right)-2\eta\right)\,,\\
 n_\perp(\eta)&=\frac{1}{2}(1-n_\parallel(\eta))\,.
\end{align}
Here, $\eta$ is the eccentricity of the spheroid. For a prolate spheroid ($R_\parallel>R_\perp$),  $\eta^2=1-\frac{R_\perp^2}{R_\parallel^2}$. 
The dielectric responses of the materials, making up the plate  and the spheroid, are modeled by the simple form
\begin{equation}
 \varepsilon_\alpha=1+\frac{C_\alpha\omega_\alpha^2}{\omega^2_\alpha-\omega^2-i\gamma_\alpha\omega}\,.
\end{equation}
The parameters are given in Table~\ref{table:1} and resemble realistic values \cite{kittel2005introduction}.
\begin{table}[t]
\begin{ruledtabular}
\begin{tabular}{cccc}
&$C_\alpha$&$\omega_\alpha$&$\gamma_\alpha$\\
\hline
Spheroid&$3$&$1\times10^{13}$&$1\times10^{11}$\\
Plate$^1$&$3$&$7\times10^{12}$&$7\times10^{10}$\\
Plate$^2$&$3$&$8.76\times10^{12}$&$8\times10^{10}$\\
\end{tabular}
\end{ruledtabular}
\caption{\label{table:1} Parameters for the oscillator model of the dielectric function of spheroid and plate, where superscripts refer to Fig.~\ref{fig:OptAnisotropie}$^1$ and Fig.~\ref{fig:OptKugelform}$^2$. $\omega_p$ and $\gamma_p$ are given in rad/sec, $C_\alpha$ is dimensionless.}
\end{table}

\begin{figure}[h]
\centering
 \includegraphics[width=\linewidth]{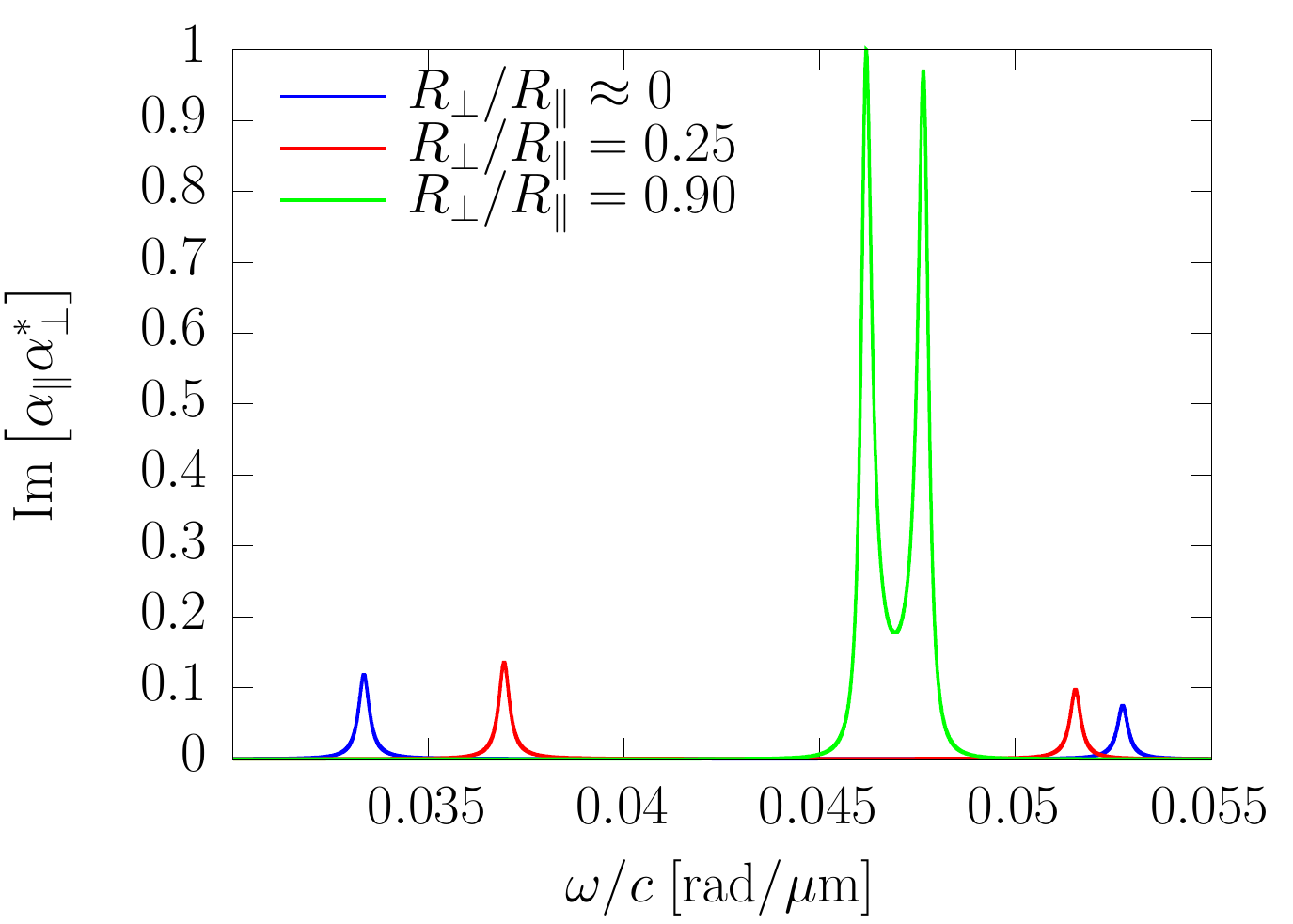}
 \caption{\label{fig:Polarizabilites} Imaginary part of $\alpha_\parallel\alpha_\perp^*$ for different ratios of $R_\perp/R_\parallel$ upon constant volume over the rescaled frequency axis $\omega/c$ in $\mathrm{rad/\mu m}$. The curves are normalized by the maximal value shown in the graph.}
\end{figure}

In order to understand the behavior of the force as a function of the involved parameters, we first investigate the factor $\Im[\alpha_\parallel\alpha_\perp^*]$ in Eq.~\eqref{eq:Lateral_force_spheroid} as a function of frequency $\omega$ for different values of $R_\parallel/R_\perp$, see Fig.~\ref{fig:Polarizabilites} (we keep the spheroid's volume fixed). 
 By rewriting the imaginary part, 
\begin{equation}
\Im\left[\alpha_\parallel\alpha_\perp^*\right]=\Re[\alpha_\perp]\Im[\alpha_\parallel]-\Re[\alpha_\parallel]\Im[\alpha_\perp]\,,
\end{equation}
we note that this term yields two contributions being responsible for the two peaks seen in Fig.~\ref{fig:Polarizabilites}. For small ratios, $R_\perp/R_\parallel\approx 0$, these two peaks are far away from each other on the frequency axis (since the resonances in $\alpha_\perp$ and $\alpha_\parallel$ are far away from each other) and  also rather small (since the products $\Re[\alpha_\perp]\Im[\alpha_\parallel]$ are small for the same reason), cf.~blue curve. For higher ratios of $R_\perp/R_\parallel$ the two contributions become larger and larger (cf.~green curve) as now the overlap between $\alpha_\perp$ and $\alpha_\parallel$ increases.  Ultimately, for $R_\perp/R_\parallel\rightarrow 1$, where the spheroid approaches a sphere, the peaks decrease to zero, as $\Im\left[\alpha_\parallel\alpha_\perp^*\right]\rightarrow 0$ for $\alpha_\parallel \rightarrow \alpha_\perp$.

\begin{figure}[t]
\centering
 \includegraphics[width=\linewidth]{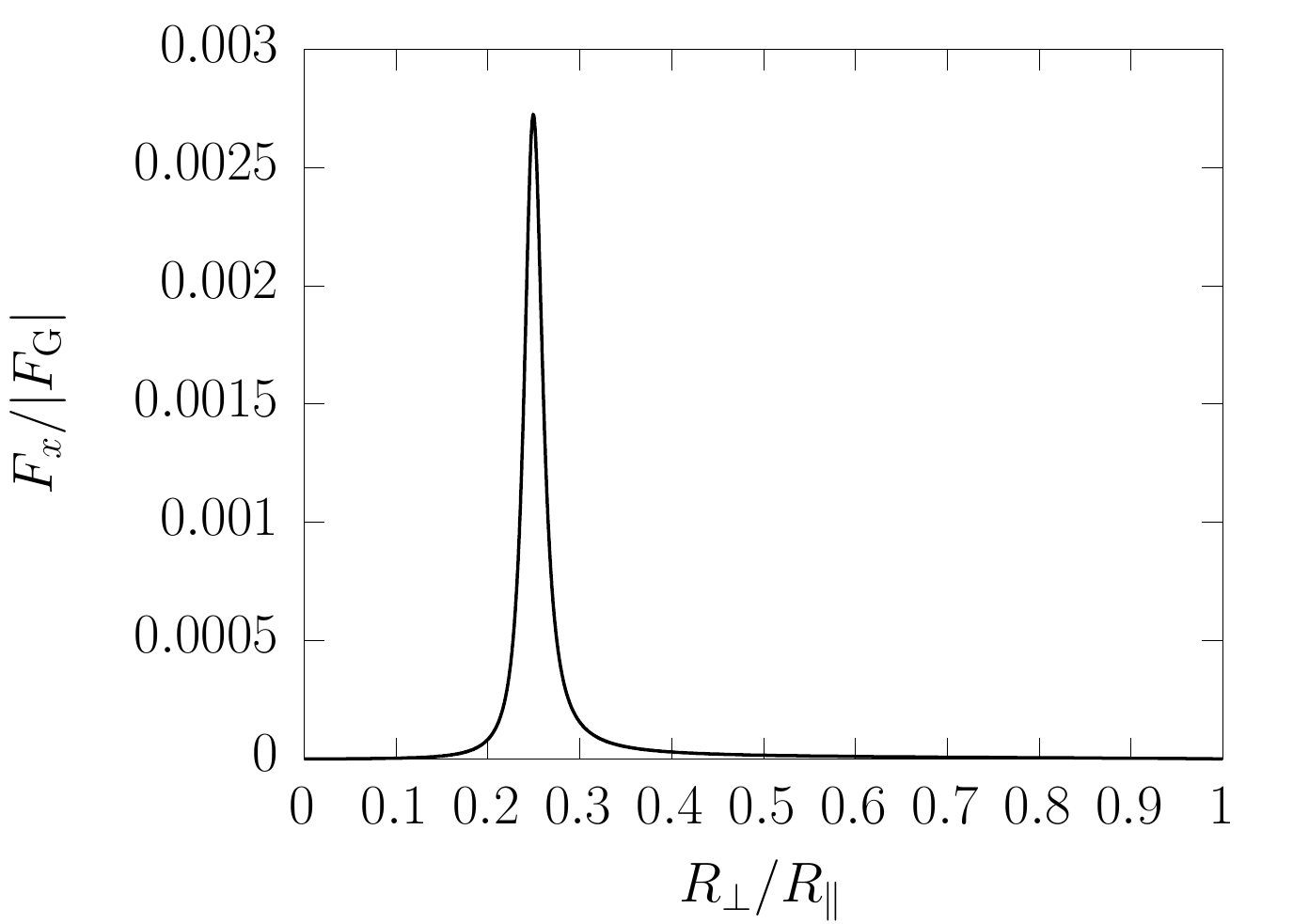}
 \caption{\label{fig:OptAnisotropie} Lateral Casimir force for a prolate spheroid (temperature $\unit[550]{K}$) of fixed radius $R_\parallel=\unit[40]{nm}$ in front of a  plate (temperature $\unit[300]{K}$) as a function of $R_\perp/R_\parallel$.  The force is normalized by the gravitation force~$|F_G|$. The distance between spheroid and plate was set at $d=\unit[400]{nm}$. The perpendicular radius $R_\perp$ was varied from $\unit[0]{nm}$ to $\unit[40]{nm}$.}
\end{figure}

Having analyzed $\Im[\alpha_\parallel\alpha_\perp^*]$, it is evident, that the factor $\Im[\frac{\varepsilon_p-1}{\varepsilon_p+1}]^2$ in Eq.~\eqref{eq:Lateral_force_spheroid}, being  peaked as a function of $\omega$ as well, strongly influences the expected force, too. Fig.~\ref{fig:OptAnisotropie} shows the lateral force (the self-propulsion force) for the peak in $\Im[\frac{\varepsilon_p-1}{\varepsilon_p+1}]^2$ located at $\omega/c=\unit[0.037]{rad/\mu m}$, corresponding to the second row in Table~\ref{table:1}. We show the force for $T=550$~K and $T_p=300$~K, and the angles $\theta=\pi/4$ and $\phi=0$ are chosen to maximize Eq.~\eqref{eq:Lateral_force_spheroid}. In order to give the force in useful units, we show its ratio to the corresponding gravitational force acting on the object, with a mass density of $\unit[3.21]{g\, cm^{-3}}$. We see that in Fig.~\ref{fig:OptAnisotropie}, the force is maximal for $R_\perp/R_\parallel\approx 0.25$, where the left peak of the red curve in Fig.~\ref{fig:Polarizabilites} overlaps with the peak of $\Im[\frac{\varepsilon_p-1}{\varepsilon_p+1}]^2$. The force is in the range of permills of the gravitational force.

In Fig.~\ref{fig:OptAnisotropie}, we show the force for the same parameters, but now $\Im[\frac{\varepsilon_p-1}{\varepsilon_p+1}]^2$ has a peak at $\omega/c=\unit[0.046]{rad/\mu m}$, corresponding to the third row in Table~\ref{table:1}. The force is now maximal for  $R_\perp/R_\parallel\approx 0.9$, where the left peak of the green curve in Fig.~\ref{fig:Polarizabilites} overlaps with the peak of $\Im[\frac{\varepsilon_p-1}{\varepsilon_p+1}]^2$. Notably, it is of the order of the gravitational force, hence being well in the detectable regime, e.g. using fly-by experiments, where the changes in particle trajectory can be detected \cite{chen2002demonstration}. Imagining a spheroid rotating around the $y$-axis while flying parallel to a surface, the periodic accelerations should be visible.
\smallskip
\begin{figure}[t]
\centering
 \includegraphics[width=\linewidth]{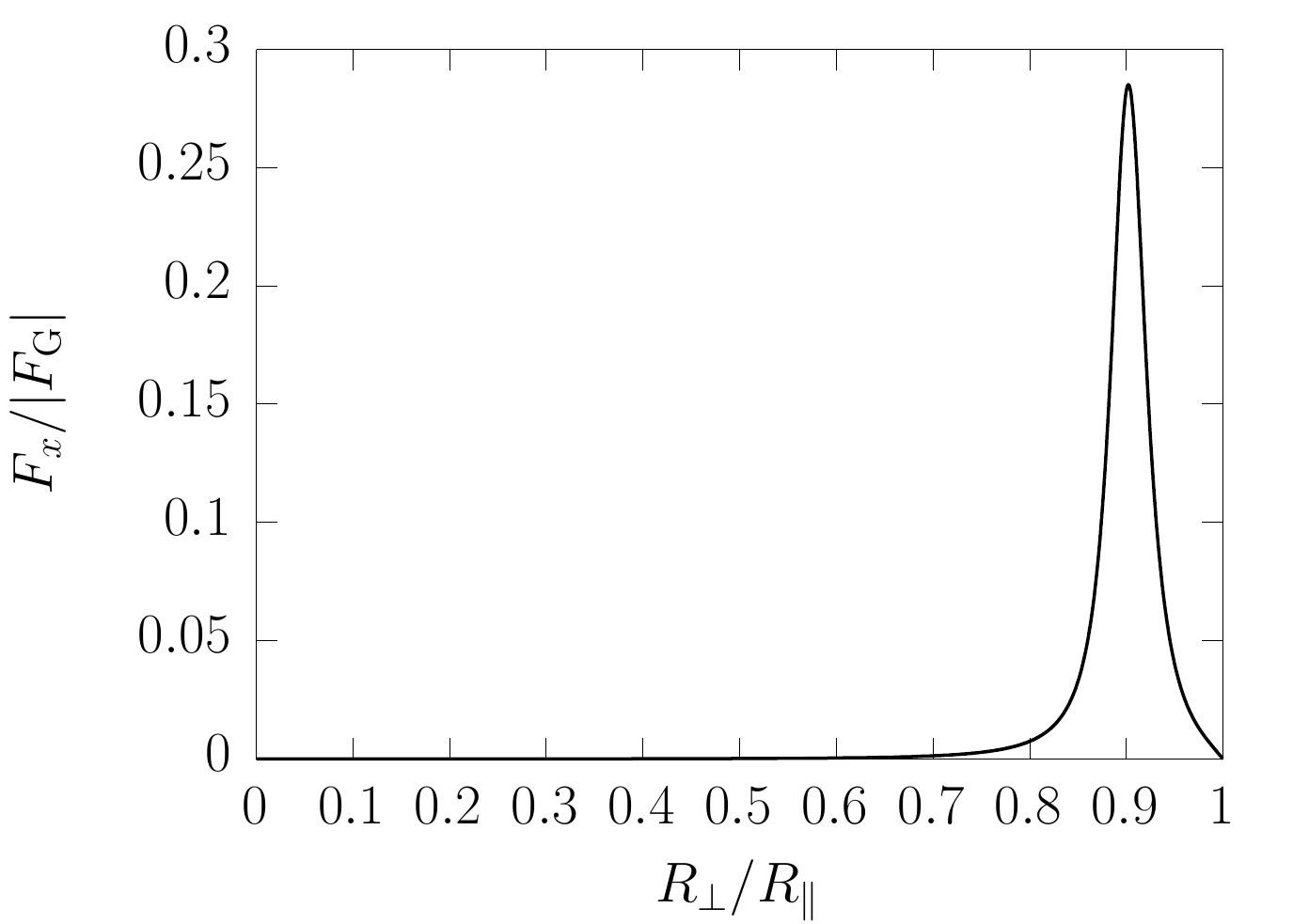}
 \caption{\label{fig:OptKugelform} Lateral Casimir force for a prolate spheroid of fixed radius $R_\parallel=\unit[40]{nm}$ in front of a  plate as a function of $R_\perp/R_\parallel$ with the same set-up parameters as in Fig.~\ref{fig:OptAnisotropie}, but with a different material of the plate (cf.~Table~\ref{table:1}).}
\end{figure}

\section{Linear response theory and Casimir friction}\label{sec:Linear_Response}
In this section, we would like to connect our findings for the self-propulsion force to other experimentally measurable quantities. In subsection \ref{sec:On}, we exploit the Onsager Theorem in order to predict the heating of the particle when flying parallel to a planar surface with a given velocity $v_x$. In subsection \ref{sec:Ts}, we discuss a subtle additional contribution to the Casimir friction force felt by the particle.
\subsection{Onsager theorem}\label{sec:On}
The Onsager Theorem (proven explicitly for fluctuational electrodynamics in Ref.~\cite{Vlad_linear_response}), relates two distinct (experimental) set-ups, where the global equilibrium is disturbed by different means. In set-up i), starting from equilibrium,  one perturbs the temperature of one of the objects, and asks for the change of Casimir force due to that. In set-up ii), again starting from equilibrium, one perturbs the system by driving one of the objects with a constant velocity, and asks for the change of its temperature (more precisely the change in its total heat absorption $H$). The Onsager Theorem then states, 
\begin{equation}\label{eq:Onsager_theorem}
 \frac{\mathrm{d}\langle H^{(\alpha)}\rangle}{\mathrm{d}\vct{v}_\beta}\bigg\vert_{\vct{v}_\beta=0}=-T\frac{\mathrm{d}\langle\vct{F}^{(\beta)}\rangle}{\mathrm{d}T_\alpha}\bigg\vert_{\{T_\alpha\}=T_\mathrm{env}=T}\,.
\end{equation}
In this equation $\alpha$ and $\beta$ can be either of the two objects $\{\alpha,\beta\}=1,2$. 

Specifically exploiting this relation for the set-up considered in the previous sections, we can predict the heating of the anisotropic particle, when it is driven parallel to the surface at constant velocity $v_x$. It follows directly from Eq.~\eqref{eq:Near_field_limit_T} (which we also have to expand in temperature differences in order to make linear response theory valid), 
	\begin{align}\label{eq:Hs}
		  &\lim_{\lambda_T\gg d\gg R}  \frac{\mathrm{d}\langle H\rangle}{\mathrm{d}v_{x}}\bigg\vert_{v_{x}=0}\hspace*{-0.045\linewidth}=\frac{27}{128\sqrt{2}\pi}\int_0^\infty\hspace*{-0.035\linewidth}\mathrm{d}\omega \frac{\mathrm{e}^{\frac{\hbar\omega}{k_B T}}}{\left(\mathrm{e}^{\frac{\hbar\omega}{k_B T}}-1\right)^2}\frac{\hbar\omega}{k_B T}\nonumber\times\\
		  &\frac{c^6}{\omega^6}\Im\left[\frac{\varepsilon_p-1}{\varepsilon_p+1}\right]^2\sum_m A_m\Re\left[\mathcal{T}_{-1,m}\mathcal{T}_{0,m}^*-\mathcal{T}_{0,m}\mathcal{T}_{1,m}^*\right]\,.
	  \end{align}
  We note that the same manipulations performed for the force in the previous sections can also be performed here, such that Eqs.~\eqref{eq:Lateral_force_general} and \eqref{eq:Lateral_force_spheroid} are as well linked to the heating of the polarizable object.

  We also note that $\frac{\mathrm{d}\langle H\rangle}{\mathrm{d}v_{x}}\vert_{v_{x}=0}$ can be positive or negative, depending on the direction of driving with respect to the orientation of the particle, so that the anisotropic particle is heated or cooled. Last, for an isotropic object, this expression is identically zero as is the self-propulsion force, and $\frac{\mathrm{d}\langle H\rangle}{\mathrm{d}v_{x}}\vert_{v_{x}=0}=0$, such that the heating of the particle will be of higher order in $v_x$.  

  \subsection{Additional Casimir friction}\label{sec:Ts}
  Imagine the anisotropic particle moving parallel to the surface with constant velocity $v_x$. Due to this motion and Eq.~\eqref{eq:Hs}, the temperature of the particle will change, with $\Delta T$ linear in $v_x$. According to our computations in the main text, this change in temperature will result in a lateral Casimir force, which constitutes an {\it additional contribution to the particle's friction} force. This friction force is linear in $v_x$, and hence additional to the standard approaches for computation of friction, as e.g. in Refs.~\cite{pieplow2013fully,RevModPhys.79.1291,Vlad_linear_response}. It appears for anisotropic particles, and regardless of the particle heating up or cooling down,  always increases the friction. From Eq.~\eqref{eq:Onsager_theorem}, this additional friction force seems to vanish for $T\to0$.
  \subsection{Time scales of force fluctuations}%\label{sec:Ts}
  Another well-known relation in linear response relates the above mentioned friction coefficient $\gamma$ to the fluctuations of forces $F(t)$ in equilibrium (the so-called Kirkwood formula \cite{kirkwood1946third,kubo1966fluctuation,volokitin2007near})
  \begin{equation}\label{eq:f}
	  \gamma(t)=\frac{1}{k_B T}\int_0^t\mathrm{d}t' \langle F(t')F(0)\rangle^\mathrm{eq}\,.
  \end{equation}
  where $\delta F(t)=F(t)-\langle F(t)\rangle^\mathrm{eq}$ represents the force fluctuations. Eq.~\eqref{eq:f} gives the time dependent friction coefficient, if, for $t<0$, the particle was at rest, and for $t\geq 0$, the particle moves with constant velocity $v$. We can now discuss two distinct time scales of that friction, following our discussion in the previous subsection. First, there is a quick adjustment of the friction due to the reaction rate of the charge and current fluctuations in the materials, which we expect on the time scale related to typical frequencies appearing in Casimir integrals, i.e, of the order of $\tau_1=10^{-14}s$. Second, as discussed above, the particle --if anisotropic-- will change its temperature, giving rise to the additional friction contribution. The time scale for it is much longer, related to the time necessary to change the particle's temperature. 

\begin{figure}[t]
\centering
 \includegraphics[width=\linewidth]{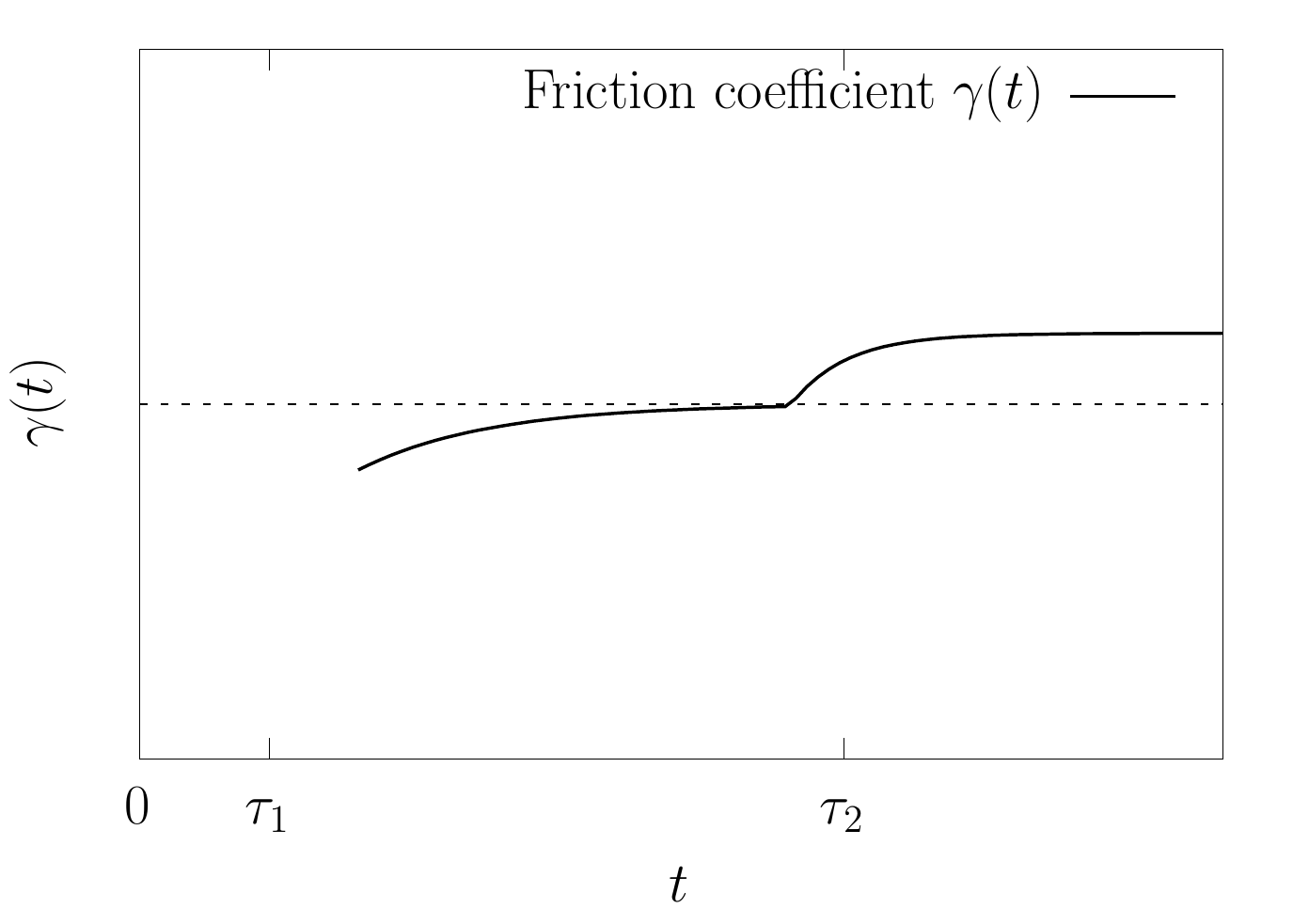}
 \caption{\label{fig:Friction_force} Principle behavior of the time-dependent friction coefficient for an anisotropic particle in front of a plate, moving at constant velocity $v_x$ for $t\geq0$. In the evolution of time two contributions show up, exhibiting the signatures of different time scales $\tau_1$ and $\tau_2$. $\tau_1$ is due to charge fluctuations, while $\tau_2$ is due to fluctuations of internal energy of the particle, and is absent for isotropic particles. The dashed line represents the stationary friction coefficient as proposed by standard approaches for computation of friction. The details for $t\to0$ are omitted due to unknown behavior in this regime.}
\end{figure}

  Interestingly, with Eq.~\eqref{eq:f}, this time scale should also visible in the fluctuations of $F(t)$, more specifically, in its time dependent autocorrelator. This is physical, as now, the force depends on the particles temperature, and therefore $\langle F(t')F(0)\rangle^\mathrm{eq}$ is influenced by temperature fluctuations of the object (or more precisely, by fluctuations of its internal energy $\cal E$). Following arguments of Ref.~\cite{Vlad_linear_response}, the time scale of fluctuations of ${\cal E}(t)$ is estimated to be $\tau_2=C/k$, where $C$ is the object's heat capacity and $k$ is the heat transfer coefficient, connecting the object to its environment (including the plate). This time scale is typically in the micro- to millisecond regime \cite{Long_paper, Vlad_linear_response} and hence much larger then the 'electronic' $\tau_1$. This discussion is sketched in Fig.~\ref{fig:Friction_force}.  

\section{Summary}
Using fluctuational electrodynamics, we have studied the phenomenon of self-propulsion through Casimir forces in thermal non-equilibrium in two scenarios. 
For an isolated object the self-propulsion force is found to be of order $\mathcal{O}(R^7)$ for small $R$ (denoting $R$ as the largest dimension of the object). As expected, the self-propulsion force vanishes for isotropic objects. We then add a planar surface to our set-up focusing on the self-propulsion of the object alongside the surface, i.e., the lateral Casimir force. In the near-field, where the separation $d$ between particle and plate is much smaller than the thermal wavelength $\lambda_T=\hbar c/k_B T$ (which is roughly $\unit[8]{\mu m}$ at room temperature), the leading order term for small and anisotropic particles is of order $\mathcal{O}(R^6)$ and diverges like $d^{-7}$. Rewriting the force formula in terms of polarizabilities, we expose the strong dependence of the lateral force on the spatial orientation of the object. For the case of a spheroid we show that the force can be as large as the gravitational force, thus being potentially measurable in experiments. 

Finally, we link our results to relations of linear response theory in fluctuational electrodynamics. Exploiting the Onsager theorem we can predict the heating or cooling of the anisotropic particle, when it is driven parallel to the surface at constant velocity $v_x$. This change in temperature leads to a lateral Casimir force, which constitutes an additional contribution to the particle's friction force linear in $v_x$ compared to the standard approaches for computation of friction. We close by specifying the time scales $\tau_1$ and $\tau_2$ of the two contributions to the friction, where the time scale of the standard approach is assumed to be of the order of $\tau_1=\unit[10^{-14}]{s}$ and the time scale of the correction term is typically in the micro- to millisecond regime.

\begin{acknowledgments}
We thank M.T.H.~Reid, E.D.~Tomlinson, G.~Bimonte,  T.~Emig, R.~L.~Jaffe, M. Kardar and N. Graham for discussions. This research was supported by Deutsche Forschungsgemeinschaft (DFG) grant No. KR 3844/2-1 and MIT-Germany Seed Fund grant No. 2746830.
\end{acknowledgments}
\begin{appendix}
\section{Green's function $\mathbb{G}$}\label{app:Greens_function}
In classical electrodynamics the electric field obeys the Helmholtz equation \cite{Jackson}
\begin{equation}\label{eq:Helmholtz}
\left[\mathbb{H}_0-\mathbb{V}-\frac{\omega^2}{c^2}\mathbb{I}\right]\vct{E}(\omega,\vct{r})=0\;,
\end{equation}
where $\mathbb{H}_0=\nabla\times\nabla$ describes free space, and 
\begin{equation}
 \mathbb{V}=\frac{\omega^2}{c^2}(\bbeps-\mathbb{I})+\nabla\times \left(\mathbb{I}-\frac{1}{\bbmu}\right)\nabla\times
\end{equation}
is the potential introduced by the objects.
Thus, the dyadic free Green's function is defined by \cite{Jackson,Partial_wave_Expansion}
\begin{equation}
 \left[\mathbb{H}_0-\mathbb{V}-\frac{\omega^2}{c^2}\mathbb{I}\right]\mathbb{G}(\vct{r},\vct{r}')=\mathbb{I}\delta^{(3)}(\vct{r}-\vct{r}')\;.
\end{equation}
Accordingly, the free Green's function $\mathbb{G}_0$ solves the wave equation for $\mathbb{V}=0$.

\section{Classical scattering operator $\mathbb{T}$}\label{app:Scattering_operator}
The classical scattering operator $\mathbb{T}$ is a convenient way of rewriting the Helmholtz equation as a Lippmann-Schwinger equation \cite{Lippmann}.
The Lippmann-Schwinger equation 
\begin{equation}
 \vct{E}^\mathrm{sc}=\vct{E}+\mathbb{G}_0\mathbb{V}\vct{E}^\mathrm{sc}
\end{equation}
is the general solution to the Helmholtz equation~\eqref{eq:Helmholtz}. Here $\mathbb{G}_0$ is the free Green's function as discussed in Appendix~\ref{app:Greens_function} and the homogeneous solution $\vct{E}$ obeys the free Helmholtz equation $\left[\mathbb{H}_0-\frac{\omega^2}{c^2}\mathbb{I}\right]\vct{E}=0$. The iterative substitution for $\vct{E}$ yields the following formal expression in terms of the $\mathbb{T}$ operator:
\begin{equation}
 \vct{E}^\mathrm{sc}=\vct{E}+\mathbb{G}_0\mathbb{T}\vct{E}\;.
\end{equation}
Solving for $\mathbb{T}$, we obtain
\begin{equation}
 \mathbb{T}=\mathbb{V}\frac{1}{1-\mathbb{G}_0\mathbb{V}}\;.
\end{equation}
The $\mathbb{T}$ operator contains all the geometric information of our objects. To the lowest order of expansion it equals the potential $\mathbb{V}$, as in the Born approximation.

\section{Spherical Basis}\label{app:Spherical_Basis}
The spherical wave basis offers the most concise representation of partial waves as it does not contain evanescent modes. Here, we adopt the wave expansion given in Ref.~\cite{Long_paper}, where the waves, depending on spherical coordinates $r$, $\theta$, and $\phi$, are defined as
\begin{align}
\vct{E}^\mathrm{reg}_{Mlm}&=\sqrt{\frac{(-1)^m\omega}{c}}\frac{1}{\sqrt{l(l+1)}}j_l\left(\frac{\omega}{c}r\right)\nabla \times \vct{r}Y^m_l(\theta,\phi)\,,\\
\vct{E}^\mathrm{out}_{Mlm}&=\sqrt{\frac{(-1)^m\omega}{c}}\frac{1}{\sqrt{l(l+1)}}h_l\left(\frac{\omega}{c}r\right)\nabla\times\vct{r}Y^m_l(\theta,\phi)\,,\\
\vct{E}^\mathrm{reg}_{Nlm}&=\frac{c}{\omega}\nabla\times\vct{E}^\mathrm{reg}_{Mlm}\,,\\
\vct{E}^\mathrm{out}_{Nlm}&=\frac{c}{\omega}\nabla\times\vct{E}^\mathrm{out}_{Mlm}\,.
\end{align}
The function $j_l$ denotes the spherical Bessel function of order $l$ and $h_l$ represents the spherical Hankel function of the first kind of order $l$. $Y_l^m(\theta, \phi)$ are the spherical harmonics, where the standard definition according to Ref.~\cite{Jackson} has to be applied. The partial wave indices in the spherical basis are given by $\mu=\{P,l,m\}$, containing the polarization $P$ (magnetic $M$ or electric $N$), the spherical multipole order $l$, as well as the multipole index $m$. Sums over partial wave indices $\Sigma_\mu$ turn into $\sum_P\sum_{l=1}^\infty\sum_{m=-l}^l$ in the spherical basis.

\section{Plane wave basis}\label{app:Plane_Wave_Basis}
The plane wave basis is a convenient way to describe planar bodies such as a (infinite) plate. We determine the $z$-direction as our symmetry axis for planar objects lying in the $xy$-plane. In the following, we consider two sets of eigenfunctions of the wave equation. The first one is applicable to problems involving thick slabs (with negligible transmission coefficient) and makes use of elementary left- and right-moving waves. The vector eigenfunctions in this basis for the two polarizations are defined according to Ref.~\cite{Long_paper} as
\begin{align}
\vct{M}^\pm_{\vct{k}_\perp}(\vct{x}_\perp,z)&=\frac{i}{2\sqrt{k_z}|\vct{k}_\perp|}(\hat{\vct{x}}k_y-\hat{\vct{y}}k_x)e^{i \vct{k}\cdot\vct{r}}\,,\\
\vct{N}^\pm_{\vct{k}_\perp}(\vct{x}_\perp,z)&=\frac{\frac{c}{\omega}}{2\sqrt{k_z}|\vct{k}_\perp}(\pm\hat{\vct{x}}k_xk_z\pm\hat{\vct{y}}k_yk_z+\hat{\vct{z}}k^2_\perp)e^{i\vct{k}\cdot\vct{r}}\,.
\end{align}
In these equations $\vct{x}_\perp$ and $\vct{k}_\perp$ are the spatial coordinate and the wave vector perpendicular to the symmetry axis. For the wave vector we can hence write $\vct{k}=(\vct{k}_\perp,k_z)^T$, with $k_z=\sqrt{\frac{\omega^2}{c^2}-k_\perp^2}$. The waves are denoted by partial wave indices $\mu=(j,P,\vct{k}_\perp)$ and $\sigma(\mu)=({\bar j},P,-{\bf k}_{\perp})$, where $\bar{L}=R$ and $\bar{R}=L$. In terms of these eigenfunctions, regular and outgoing waves read as
\begin{align}
\vct{E}^\mathrm{reg}_{R,\,P,\,\vct{k}_\perp}(\vct{r})&=\vct{P}^-_{\vct{k}_\perp}(\vct{x}_\perp,z)\,,\\
\vct{E}^\mathrm{reg}_{L,\,P,\,\vct{k}_\perp}(\vct{r})&=\vct{P}^+_{\vct{k}_\perp}(\vct{x}_\perp,-z)\,,\\
\vct{E}^\mathrm{out}_{R,\,P,\,\vct{k}_\perp}(\vct{r})&=\begin{cases}2\vct{E}^\mathrm{reg}_{R,\,P,\,\vct{k}_\perp}(\vct{r}), \qquad &z\ge0 \\
0, \qquad &z<0 \end{cases}\,,\label{eq:Plane_wave_eigenfunctions}\\
\vct{E}^\mathrm{out}_{L,\,P,\,\vct{k}_\perp}(\vct{r})&=\begin{cases}0, \qquad &z\ge0 \\
2\vct{E}^\mathrm{reg}_{L,\,P,\,\vct{k}_\perp}(\vct{r}), \qquad &z<0 \end{cases}\,.
\end{align}
The second set of eigenfunctions employs waves of definite parity that are convenient in problems involving slabs of finite thickness whose transmission coefficient cannot be neglected. These waves have a definite parity (carrying the index $s=\pm$) under reflections at the $z=0$ plane and consist of a superposition of left- and right-moving waves:
\begin{align}
\vct{E}^\mathrm{reg}_{s,P,\vct{k}_\perp}(\vct{r})=\frac{i^{\frac{1-s}{2}}}{\sqrt{2}}[\vct{E}^\mathrm{reg}_{R,\,P,\,\vct{k}_\perp}(\vct{r})+s\vct{E}^\mathrm{reg}_{L,\,P,\,\vct{k}_\perp}(\vct{r})]\,,\\
\vct{E}^\mathrm{out}_{s,P,\vct{k}_\perp}(\vct{r})=\frac{i^{\frac{1-s}{2}}}{\sqrt{2}}[\vct{E}^\mathrm{out}_{R,\,P,\,\vct{k}_\perp}(\vct{r})+s\vct{E}^\mathrm{out}_{L,\,P,\,\vct{k}_\perp}(\vct{r})]\,.\label{eq:Transformation_identity_moving_waves}
\end{align}
The waves carry the partial wave indices $\mu=(s,P,\vct{k}_\perp)$ and $\sigma(\mu)=(s,P,-\vct{k}_\perp)$. The transition between the two sets of outgoing waves can be written in terms of the unitary transformation
\begin{equation}
\vct{E}^\mathrm{out}_{s,P,\vct{k}_\perp}=\sum_{j=L,R}\vct{E}^\mathrm{out}_{j,P,\vct{k}_\perp} O_{js}\,.
\end{equation}
The transformation matrix immediately follows from Eq.~\eqref{eq:Transformation_identity_moving_waves} and is given by
\begin{equation}\label{eq:Transformation_Matrix_Plane_Waves}
O_{js}=\frac{1}{\sqrt{2}}(\delta_{s,+}\delta_{j,R}+\delta_{s,+}\delta_{j,L}+i\delta_{s,-}\delta_{j,R}-i\delta_{s,-}\delta_{j,L})\,.
\end{equation}
\section{$\mathcal{T}$ matrix}\label{app:T_Elements}
\subsection{General definition}
The matrix elements of the classical scattering operator $\mathbb{T}$ are readily evaluated as \cite{Long_paper}
\begin{equation}\label{eq:T_Matrix_Elements}
\mathcal{T}_{\mu\mu'}=i\int\mathrm{d}^3\vct{r}\int\mathrm{d}^3\vct{r}'\vct{E}^\mathrm{reg}_{\sigma(\mu)}(\vct{r})\mathbb{T}(\vct{r},\vct{r}')\vct{E}^\mathrm{reg}_{\mu'}(\vct{r}')\,.
\end{equation}
The function $\sigma(\mu)$ is a permutation among the partial wave indices, which fulfills $\sigma(\sigma(\mu))=\mu$. In the spherical basis $\sigma(\mu)$ changes the multipole index $m$ to $-m$, i.e.~$\sigma(\mu)~=~\{P,l,-m\}$. Note that the matrix elements satisfy the condition $\mathcal{T}_{\mu\mu'}=\mathcal{T}_{\sigma(\mu')\sigma(\mu)}$ because of the symmetry of $\mathbb{T}$.

\subsection{$\mathcal{T}$ matrix of a sphere}
The scattering of electromagnetic waves by a sphere of radius $R$ and permittivity $\varepsilon$ is often referred to as Mie scattering. In this specific case the boundary value problem can be solved analytically and one obtains the matrix elements of $\mathcal{T}$ in closed form, the so-called Mie coefficients. These coefficients are well-known quantities and thoroughly defined in Ref.~\cite{Tsang_Kong_Ding}. Assuming isotropic and local $\varepsilon$ and $\mu$ the $\mathcal{T}$ matrix is diagonal in all indices and most notably independent of $m$, $\mathcal{T}^{PP'}_{lml'm'}~=~\mathcal{T}_l^P\delta_{PP'}\delta_{ll'}\delta_{mm'}$. Introducing the abbreviations $R^*=R\omega/c$ and $\tilde{R}^*=\sqrt{\varepsilon\mu}R\omega/c$, the matrix element $\mathcal{T}^P_l$ can be written as \cite{Tsang_Kong_Ding}
\begin{equation}\label{eq:Tm}
 \mathcal{T}_{l}^M=-\frac{\mu j_l(\tilde R^*)\frac{d}{d R^*}\left[R^*j_l(R^*)\right]-j_l(R^*)\frac{d}{d \tilde R^*}[\tilde R^*j_l(\tilde R^*)]}{\mu j_l(\tilde R^*)\frac{d}{d R^*}\left[R^*h_l(R^*)\right]-h_l(R^*)\frac{d}{d \tilde R^*}[\tilde R^*j_l(\tilde R^*)]}\,.
\end{equation}
$\mathcal{T}^N_l$ is obtained from $\mathcal{T}^M_l$ by interchanging $\mu$ and $\varepsilon$.

\subsection{$\mathcal{T}$ matrix of a plate}
The $\mathcal{T}$ matrix of a plane-parallel dielectric slab of finite thickness in the region $-l\le z\le 0$ is closely related to the Fresnel reflection and transmission coefficients $r_P^{(R)}$ and $t_P^{(R)}$ for outgoing waves on the right of the slab by \cite{Long_paper}
\begin{align}
{\tilde {\cal T}}^{P,P'}_{R,{\bf k}_{\perp},R,{\bf k}'_{\perp}}&=\delta_{PP'}(2 \pi)^2 \delta^{(2)}({\bf k}_{\perp}-{\bf k}'_{\perp})\; \frac{ t^{(R)}_P -1}{2}\,,\\
{\tilde {\cal T}}^{P,P'}_{R,{\bf k}_{\perp},L,{\bf k}'_{\perp}}&=  \delta_{PP'}(2 \pi)^2 \delta^{(2)}({\bf k}_{\perp}-{\bf k}'_{\perp}) \;\frac{r^{(R)}_P}{2} \,.
\end{align}
By substituting $L$ for $R$ we obtain the equivalent matrix elements for outgoing waves on the left side of the slab.
Here, we provide a slightly different variant of the scattering matrix, the transformed matrix $\tilde{\mathcal{T}}$, which is linked to the original $\mathcal{T}$ matrix by application of the unitary transformation $O$ defined in Eq.~\eqref{eq:Transformation_Matrix_Plane_Waves} reading as $\tilde{\mathcal{T}}=O\mathcal{T}O^\dagger$. The Fresnel coefficients both depend on the thickness $l$ of the slab (c.f. Ref.~\cite{landau1984electrodynamics}, p.299). Considering an infinitely thick plate ($l\rightarrow\infty$), the transmission vanishes and $r_P^{(R)}$ approaches the relation given in Ref.~\cite{Jackson}:
\begin{align}\label{eq:Fresnel_coeff}
r^N\left(k_\perp,\omega\right)=\frac{\varepsilon\sqrt{\frac{\omega^2}{c^2}-k_\perp^2}-\sqrt{\varepsilon\mu\frac{\omega^2}{c^2}-k_\perp^2}}{\varepsilon\sqrt{\frac{\omega^2}{c^2}-k_\perp^2}+\sqrt{\varepsilon\mu\frac{\omega^2}{c^2}-k_\perp^2}}\,.
\end{align}
The Fresnel reflection coefficient for magnetic polarization $r^M$ is obtained from $r^N$ by interchanging $\varepsilon$ and $\mu$.

\section{Conversion matrices}\label{app:Conversion_Matrix}
\subsection{Plane waves to spherical waves}
The outgoing plane wave eigenfunctions in Eq.~\eqref{eq:Plane_wave_eigenfunctions} are expanded in regular spherical waves by means of the conversion matrix $D$:
\begin{equation}
\vct{E}^\mathrm{out}_{R,P,\vct{k}_\perp}(\vct{x}_\perp,z)=\sum_{P',l,m}D_{lmPP'\vct{k}_\perp}\vct{E}^\mathrm{reg}_{P'lm}\,.
\end{equation}
The matrix elements of $D$ are given according to Ref.~\cite{Long_paper} by
\begin{subequations}
 \begin{align}
D_{lmMM\vct{k}_\perp}=&\frac{-i^{l+1}}{\sqrt{(-1)^m}}\frac{c}{\omega}\sqrt{\frac{4\pi(2l+1)(l-m)!}{l(l+1)(l+m)!}}\nonumber\times\\
&\sqrt{\frac{c}{\omega}}\frac{|\vct{k}_\perp|}{\sqrt{k_z}}  P^{'m}_l\bigg(\frac{c}{\omega}\sqrt{\frac{\omega^2}{c^2}-k_\perp^2}\bigg)e^{-i m\Phi_{\vct{k}_\perp}}\,,\\
D_{lmNM\vct{k}_\perp}=&\frac{mi^{l+1}}{\sqrt{(-1)^m}}\sqrt{\frac{4\pi(2l+1)(l-m)!}{l(l+1)(l+m)!}}\nonumber\times\\
& \frac{1}{|\vct{k}_\perp|} P^m_l\bigg(\frac{c}{\omega}\sqrt{\frac{\omega^2}{c^2}-k^2_\perp}\bigg)e^{-i m \Phi_{\vct{k}_\perp}}\,,\\
D_{lmNN\vct{k}_\perp}&=D_{lmMM\vct{k}_\perp}\,,\\
D_{lmMN\vct{k}_\perp}&=D_{lmNM\vct{k}_\perp}\,,
\end{align}
\end{subequations}
where $\Phi_{\vct{k}_\perp}$ is the angle of $\vct{k}_\perp$ with respect to the $x$ axis.

\subsection{Spherical waves to spherical waves}
Outgoing spherical waves can be expanded into regular spherical waves with respect to a different origin by application of the translation matrix $\mathcal{U}$. The expansion of outgoing waves of object 1 in the coordinate system of object 2 can be written as
\begin{equation}
\vct{E}^\mathrm{out}_{Plm}(\vct{r})=\sum_{P'l'm'}\mathcal{U}^{21}_{P'P,l'l,m'm}(d)\vct{E}^\mathrm{reg}_{P'l'm'}(\vct{r}+d\hat{\vct{z}})\,.
\end{equation}
The translation matrix $\mathcal{U}^{21}$ implies a shift of waves along the $z$ axis by length $d$, where $d\hat{\vct{z}}=\mathcal{O}_1-\mathcal{O}_2$ is the connection vector between the origins of the coordinate system of object 1 and 2.
The matrix elements of $\mathcal{U}$ are readily calculated according to Ref.~\cite{Wittmann}:
\begin{align}
\mathcal{U}_{P'P,l'l,m'm}=&\sum_\nu\bigg[\frac{l(l+1)+l'(l'+1)-\nu(\nu+1)}{2}\delta_{PP'}\nonumber\\
&-i m d\frac{\omega}{c}(1-\delta_{PP'})\bigg]A_{l'l\nu m}(d)\delta_{mm'}\,.
\end{align}
The function $A_{l'l\nu m}(d)$ is defined as
\begin{align}
A_{l'l\nu m}(d)&=(-1)^m\,i^{l-l'+\nu}(2\nu+1)\sqrt{\frac{(2l+1)(2l'+1)}{l(l+1)l'(l'+1)}}\nonumber\times\\
&\begin{pmatrix} l & l' & \nu\\0 & 0 &0 \end{pmatrix}\begin{pmatrix} l & l' & \nu\\m & -m &0 \end{pmatrix}h_\nu\left(\frac{d\omega}{c}\right)\,,
\end{align}
with $h_\nu$ the spherical Hankel function of the first kind. By replacing $h_\nu$ with $j_\nu$, the spherical Bessel function of the first kind, we obtain the regular part of the translation matrix that is linked to the infinitesimal translation operator $\vct{p}$ by
\begin{equation}
 \vct{p}_{\mu\mu'}=-\nabla_a\mathcal{V}_{\mu\mu'}(\vct{a})\vert_{\vct{a}=\vct{0}}\,.
\end{equation}

\section{Polarizability tensor $\hat{\alpha}$}\label{app:Polarizability_Tensor}
In the dipole limit $l=1$ for electrical polarization $P=N$, the $\mathcal{T}$ matrix elements for objects featuring a rotational axis of symmetry can be rewritten in terms of the (anisotropic) polarizability tensor $\hat{\alpha}$ as
\begin{equation}\label{eq:T_Matrix_Elements_Dipole_Limit}
\mathcal{T}^{N,N}_{1m,1m'}=i\frac{\omega^2}{c^2}\int\mathrm{d}\Omega\;\mathbf{E}_{N,1,-m}^{\mathrm{Reg}}(r=0)\,\hat{\alpha}\, \mathbf{E}_{N,1,m'}^{\mathrm{Reg}}(r=0)\,,
\end{equation}
where we have to perform an integration over the solid angle $\Omega$. The polarizability tensor contains information about the electrical properties of the object as well as the orientation of the object. It can be assumed diagonal in a properly chosen coordinate system. By transformation of this diagonal tensor, 
\begin{equation}
 \hat{\alpha}_0=\begin{pmatrix}
                 \alpha_x & 0 & 0\\
                 0 & \alpha_y & 0\\
                 0 & 0 & \alpha_z
                \end{pmatrix}\,,
\end{equation}
we obtain the polarizability tensor in the global frame by a passive rotation of the coordinate system around the $y$- and $z$-axis, respectively, as
\begin{equation}
\hat{\alpha}\equiv\begin{pmatrix}
                 \alpha_{xx} &  \alpha_{xy} &  \alpha_{xz}\\
                  \alpha_{yx} & \alpha_{yy} &  \alpha_{yz}\\
                  \alpha_{zx} &  \alpha_{zy} & \alpha_{zz}
                \end{pmatrix}=\hat{O}^T_z(\hat{O}^T_y\hat{\alpha}_0\hat{O}_y)\hat{O}_z \,.
\end{equation}
The rotation matrices $O_{\{y,z\}}$ are given for example in Ref.~\cite{bronshtein2007handbook}.
The polarizability tensor $\hat{\alpha}$ is a convenient way to calculate the $\mathcal{T}$ matrix elements used in this limit, as the computation of the $\mathbb{T}$ operator in Eq.~\eqref{eq:T_Matrix_Elements} is fairly difficult for arbitrary shape of the object. Note that the dipole limit is applicable only if the size of the object (denoting $R$ as the largest dimension of the anisotropic object) is the smallest scale involved, i.e., if $R$ is small compared to the thermal wavelength (which is roughly $\unit[8]{\mu m}$ at room temperature), as well as the material skin depth. In this case, the $\mathcal{T}^{N,N}_{1m,1m'}$ elements as defined in Eq.~\eqref{eq:T_Matrix_Elements_Dipole_Limit} are the dominant ones.
\end{appendix}

\end{document}